\documentclass[journal=jacsat,manuscript=article]{achemso}
\usepackage[version=3]{mhchem} % Formula subscripts using \ce{}
\usepackage{graphicx}
\usepackage{dcolumn}
\usepackage{bm}
\usepackage{natbib}
\usepackage{amssymb}
\usepackage{epsfig}
\usepackage{graphics}
\usepackage{graphicx}

\author{J. C. Garcia}
\author{D. B. de Lima}
\affiliation[Escola Polit\'ecnica, Universidade de S\~ao Paulo,
CP 61548, CEP 05424-970, S\~ao Paulo, SP, Brazil]
{Universidade de S\~ao Paulo, S\~ao Paulo, SP, Brazil}
\author{L. V. C. Assali}
\affiliation[Instituto de F\'{\i}sica,Universidade de S\~ao Paulo,
CP 66318, CEP 05315-970, S\~ao Paulo, SP, Brazil]
{}
\author{J. F. Justo}
\email{jjusto@lme.usp.br}
\affiliation[Escola Polit\'ecnica, Universidade de S\~ao Paulo,
CP 61548, CEP 05424-970, S\~ao Paulo, SP, Brazil]{Universidade de S\~ao Paulo, S\~ao Paulo, SP, Brazil}

\title[graphane]
{Group-IV graphene- and graphane-like nanosheets}

\begin{document}
%%%%%%%%%%%%%%%%%%%%%%%%%%%%%%%%%%%%%%%%%%%%%%%%%%%%%%%%%%%%%%%%%%%%%
%% The manuscript does not need to include \maketitle, which is
%% executed automatically.  The document should begin with an
%% abstract, if appropriate.  If one is given and should not be, the
%% contents will be gobbled.
%%%%%%%%%%%%%%%%%%%%%%%%%%%%%%%%%%%%%%%%%%%%%%%%%%%%%%%%%%%%%%%%%%%%%

\begin{abstract}
We performed a first principles investigation on
the structural and electronic properties of group-IV (C, SiC, Si, Ge, and Sn)
graphene-like sheets in flat and buckled configurations and the
respective hydrogenated or fluorinated graphane-like ones.
The analysis on the energetics, associated with the formation of those
structures, showed that fluorinated graphane-like sheets are very stable,
and should be easily synthesized in laboratory. We also studied the changes
on the properties of the graphene-like sheets, as result of
hydrogenation or fluorination. The interatomic distances
in those graphane-like sheets are consistent with the respective crystalline
ones, a property that may facilitate integration of
those sheets within three-dimensional nanodevices.
\end{abstract}

%\pacs{73.22.-f, 81.05.Zx, 61.48.De}
%\maketitle

\section{Introduction}

The properties of graphene, the one-atom-thick sheet with carbon atoms with
the sp$^2$ hybridization, were first discussed in the literature more than
sixty years ago \cite{castro}. It has been long considered only a theoretical
curiosity of impossible experimental realization.
This perception changed radically a few years ago, after the separation of
graphene sheets with single and multiple layers \cite{novoselov,geim,wang}. Since
then, graphene has been intensively investigated, with focus on its physical
and chemical properties \cite{geim2}. This material carries unique properties
that allows to envision a number of potential applications, such as chemical
sensors \cite{pumera,casolo}, nanoelectronic devices \cite{sch}, or
hydrogen storage systems \cite{bou}.

Graphene could be considered as a prototypical material to study the properties
of other two-dimensional nanosystems. Recently, several two-dimensional
structures have been explored in the literature.
For example, graphane, a fully hydrogenated graphene sheet with all
carbon atoms in the sp$^3$ hybridization, has been
proposed by theoretical investigations \cite{sofo} and was later
synthesized \cite{elias}. Graphene-like sheets, made of silicon
carbide \cite{huda,beka}, silicon \cite{nakano,voon},
germanium \cite{caha,houssa},  boron nitride \cite{tang,tops}, and
zinc oxide \cite{zhang} have also been discussed in the literature.

Here, we performed a systematic investigation on the trends in the properties of
group-IV (C, SiC, Si, Ge, and Sn) graphene-like structures, in flat and
buckled configurations, using
first principles total energy calculations. We then observed the modifications
on those properties as result of full coverage of hydrogen and fluorine atoms,
to form sp$^3$ graphane-like structures. We found that hydrogenation and
fluorination processes provide structures that were energetically very
accessible for all compounds, and should be easily synthesized in laboratory.
We also found that all group-IV graphene-like structures present null gaps
in both flat or buckled configurations, that opened up with hydrogenation or
fluorination in most materials. The only exception was the fluorinated
graphane-like tin, that although tin atoms were fourfold coordinated, the
material presented a null gap. This paper is organized as follow, we first
discuss the methodology, then the properties of group-IV graphene-like sheets.
Finally, we discuss the energetics and resulting physical properties
of hydrogenated and fluorinated graphane-like sheets.

\section{Methodology}

The calculations were performed using the Vienna {\it ab initio} simulation
package (VASP) \cite{kresse1}. The electronic exchange-correlation potential was
described within the density functional theory and the generalized gradient
approximation (DFT-GGA) \cite{pbe}. The electronic wave-functions were
described by a projector augmented wave (PAW) method \cite{kresse2}, taking a
plane-wave basis set with an energy cutoff of 550 eV. For all calculations,
convergence in total energy was set to 0.1 meV/atom between two self-consistent
iterations. Configurational optimization was performed by considering
relaxation in all atoms, without symmetry constrains, until forces were lower
than 3 meV/\AA \ in any atom. The Brillouin zone was sampled by a
$15\times 15 \times 1$ Monkhorst-Pack k-point grid \cite{mp}. The planar
structures were built using periodic boundary conditions with a hexagonal
simulation cell. In the direction perpendicular to the sheets ($z$), we
used a lattice parameter of 20 \AA, which was large enough to prevent
image interactions.

Binding and formation energies for all systems were computed following the
same procedure presented elsewhere \cite{sofo}. The binding energy ($\rm E_B$)
of a certain structure was computed as the difference between the total energy
of that stable structure and the total energies of the respective isolated
atoms in their neutral charge states. The formation energy ($\rm E_F$) of a
certain hydrogenated (or fluorinated) sheet was computed as the difference
between the binding energy of the graphane-like structure and the binding
energies of the respective (stable) graphene-like structure and those
energies of the diatomic molecules H$_2$ (or $F_2$). In group-IV materials,
we found that the stable graphene-like structure was the buckled
configuration (lower in energy), except for carbon.

The total energies of the isolated atoms and diatomic molecules were obtained
considering a large simulation cell and the same methodological approximations
of all the other calculations described in the previous paragraphs.

To check the validity of all approximations used in this investigation, we
compared the properties of graphene with available data from experiments and
other theoretical investigations. The computed binding energy of graphene was
-7.848 eV/atom, being 0.136 eV/atom lower than the respective energy of the
diamond cubic structure. Those two values are in excellent agreement with other
investigations \cite{yin,popok}. In terms of the structural
properties of graphene, the carbon-carbon interatomic distance was 1.425 \AA,
which is in excellent agreement with the respective experimental values
(1.42 \AA) \cite{castro}, but a little larger than the one (1.414 \AA) of
a recent theoretical investigation \cite{voon}. It should be pointed out that
while our investigation used a generalized gradient approximation the other
investigation used the local density approximation, that is known
to underestimate interatomic distances \cite{voon}.

\section{Results}

Figure 1 presents a schematic representation of the graphene-like
structures in flat (labeled $\alpha$) and buckled (labeled $\beta$)
configurations and their hydrogenated and fluorinated graphane-like forms.
\ref{tab1} presents the structural properties of group-IV graphene-like
sheets and their respective binding energies.
According to the table, the graphene-like structures of
Si, Ge and Sn in flat ($\alpha$) configurations are metastable, with
the respective buckled ones ($\beta$) being energetically more favorable,
consistent with other investigations for Si \cite{caha,voon}.

Figure 2 presents the theoretical interatomic distances and binding
energies of all group-IV graphene-like and graphane-like structures as function
of the respective properties in the (diamond cubic or zinc-blende) crystalline
solid phases, in which all group-IV atoms are in the sp$^3$
hybridization \cite{solid}. According to Figure 2a, interatomic
distances between group-IV atoms in flat graphene-like structures are on
average 5 \% shorter than those distances in the respective solid phases.
These results show that the group-IV atoms, in a sp$^2$ environment, behave
essentially the same way as carbon atoms do. For the buckled configurations,
those distances are always larger than the respective ones in the flat
configurations. Buckling distances ($\Delta z$) are consistent with recent
theoretical results for buckled sheets of silicon and germanium \cite{caha}.

Figure 2b shows that the binding energies of most group-IV
graphene-like structures in either flat or buckled configurations, except
for carbon, are higher than the respective energies in the solid phases.
This indicates that graphene-like structures, with atoms in the sp$^2$
hybridization, are not very stable when compared to the respective solid
stable phases, in which atoms are in the sp$^3$ hybridization.
These results are consistent with the generally large energy difference between
those two hybridizations in most covalent materials, being small only for
carbon. Additionally, the binding energy in the buckled configurations is
larger than the one in the flat configurations, except for carbon. In the
case of carbon, the calculations indicated that the buckled configuration is
unstable, relaxing toward the flat one.
An interesting case is SiC, in which the binding energy difference between
flat and buckled configurations is only 1 meV/atom, but the
buckling is also small. All those graphene-like
structures, in either flat or buckled configurations, presented a null
electronic gap, except for SiC, that presented a large gap of 2.54 eV.
This value is in excellent agreement with a recent theoretical investigation
using similar approximations \cite{huda,beka}.

Figure 3 presents the electronic band structure of all graphene-like
structures in flat and buckled configurations. All group-IV graphene-like
structures (of C, Si, Ge, and Sn) in a flat configuration (fig. 3a)
present a similar electronic band structure, with a band crossing in the Dirac
(K) points at the Fermi level. For all of those materials, there is linear
dispersion around those Dirac points, a property that results from the
honeycomb structure. In buckled configurations (Fig. 3b),
the linear dispersion around those Dirac points is maintained.

The electronic band structures of the flat graphene-like configurations
differ among themselves only by the fact that, in structures of C and Si,
the system is semi-metallic, being metallic in Ge and Sn ones. Such difference in the band
structure could be understood by the following explanation. For graphene-like flat
structures of C and Si, there is a specific energy band that stays over the
Fermi level in all the Brillouin zone. However, for graphene-like flat
structures of Ge and Sn, the same band crosses the Fermi
level in the $\rm \Gamma \rightarrow M$ symmetry direction, and the
system is metallic. The electronic band structure of the flat and buckled
configurations differ by the fact that the later ones do not present the
band crossing described in this paragraph.

Since the electronic band structure of group-IV graphene-like materials, in
flat configurations, are equivalent to the one of carbon, we computed the
carrier velocities around their respective Dirac points. From the results of
Figure 3, the computed carrier velocities in those points are
0.91, 0.58, 0.59, and 0.52$\times$10$^6$ m/s for flat graphene-like of
C, Si, Ge, and Sn, respectively. The computed carrier
velocities in those points are 0.46, 0.69, and 0.95  $\times$10$^6$ m/s
for buckled graphene-like of Si, Ge, and Sn, respectively. Those results
indicate that carrier velocities around the Dirac points could be very large
in the buckled configurations. Our result for graphene is in good agreement
with the experimental value of 1.1$\times$10$^6$ m/s (Refs. \cite{novoselov}
and \cite{zhang2}) and with the theoretical one of
0.63$\times$10$^6$ m/s (Ref. \cite{voon}).

\ref{tab1} presents the structural parameters for
hydrogenated and fluorinated graphane-like structures and
their respective binding and formation energies. Here, we considered
only systems associated with the chair-like configurations,
and neglected the boat-like isomeric ones. This is justified by recent
theoretical investigations for graphane \cite{sofo,leena} and fluorinated
graphane \cite{charlier}, indicating that the chair-like configuration is
energetically more favorable than the boat-like one. As described in
Figure 1c (or 1d), the chair-like configuration has hydrogen
(or fluorine) atoms alternating over and below the plane containing the
group-IV atoms. Incorporation of either hydrogen or fluorine atoms leads to very
stable structures, with binding energies (per atom) for graphane-like
structures larger than the ones for
graphene-like, as shown in Figure 2b. Additionally, the
graphane-like structures have large formation
energies in most cases, consistent with other theoretical investigations
for hydrogen incorporation in graphene \cite{sofo} and
in boron nitride graphene-like structures \cite{tang}.

Figure 2b shows the trends in the binding energies (per atom)
for hydrogenated and fluorinated graphane-like structures. The
fluorinated structures are energetically more stable than the hydrogenated
ones, and become considerably favorable for Si, Ge, and Sn materials.
These results are consistent with available experimental results for graphane
and fluorinated graphane structures \cite{leena,elias}. Therefore, it is
expected that those fluorinated graphane-like forms should be
easily synthesized in laboratory.

In terms of the structural properties of hydrogenated and fluorinated forms,
\ref{tab1} presents the interatomic and buckling distances. The interatomic
distances between the group-IV atoms and hydrogen (or fluorine) atoms are in
excellent agreement with the respective distances in typical organic molecules.
For example, in graphane ($\rm C_2H_2$) the C-C, C-H, and
buckling ($\Delta z$) distances are 1.536, 1.110, and 0.459 \AA\ agree
very well with recent theoretical results\cite{voon} of 1.520, 1.084, and
0.45 \AA, respectively. For fluorinated graphane-like structure ($\rm C_2F_2$),
the C-C and C-F distances are 1.583 and 1.382 \AA \ that agree well with
recent theoretical results \cite{leena} of 1.579 and 1.371 \AA, respectively.
According to Figure 2a, along the series, interatomic distances between group-IV atoms,
in either hydrogenated or fluorinated forms, are all very close to
the interatomic distances in their respective crystalline forms.

The results indicate that group-IV atoms, in hydrogenated and fluorinated
graphane-like structures, are fourfold coordinated and have a near tetrahedral
configuration, and their interatomic distances and binding are close to
the ones in a crystalline environment. The structures deviated
from a tetrahedral configuration, evidenced by the  buckling distance ($\Delta z$),
due to some ionic character in the
binding between the group-IV atoms and the hydrogen (or fluorine) neighboring
atoms. The results suggest that hydrogenation or fluorination
may generate two-dimensional structures that could be easily incorporated in
the surface of the respective three-dimensional crystalline counterparts. Therefore, while
integration of graphene-like structures in three-dimensional devices is
still difficult, due to large lattice mismatch,
it may be easier for hydrogenated and fluorinated graphane-like structures.

According to Figures 3c and 3d, hydrogenation and
fluorination open the electronic gap of the graphene-like structures.
In all cases, electronic gap is larger in the hydrogenated configurations
than in the fluorinated ones. An interesting case is the fluorinated
graphane-like tin ($\rm Sn_2 F_2$), in which although tin atoms have a fourfold
coordination, the material has a null gap. This result indicates that carrier
velocities should be very large in this system, even with tin atoms with all
valence electrons paired with neighboring atoms.

\section{Summary}

In summary, we investigated the trends on the structural and electronic
properties of graphene-like structures made of group-IV atoms, in terms of
their energetics and electronic band structure. The results indicate that
while the graphene-like structures (of Si, Ge, Sn, and SiC) appear
to have low stability, the respective hydrogenated and fluorinated graphane-like
ones are very stable and should be easily synthesized in laboratory.

The hydrogenated and fluorinated graphane-like structures present the group-IV
atoms in a fourfold configuration and in a near tetrahedral configuration.
Interatomic distances in those configurations are close to the respective
ones in the solid phase counterparts, a property that could facilitate integration
of those two-dimensional structures within three-dimensional nanodevices.

\vspace{1cm}

\noindent
{\bf Acknowledgments:}
The authors acknowledge support from Brazilian agencies CNPq and FAPESP.

\pagebreak

\begin{table}[!h]
\caption{Structural and electronic properties of graphene-like sheets
($\alpha$-XY and $\beta$-XY respectively for flat and buckled sheets) and
hydrogenated ($\rm XY H_2$) or fluorinated ($\rm XY F_2$) graphane-like
sheets with X = Y = C, Si, Ge, or Sn (or X = C and Y = Si for SiC).
The table presents the lattice parameter ($a$), interatomic distances ($d$),
buckling distances ($\Delta z$), binding ($\rm E_{B}$), formation ($\rm E_{F}$),
and electronic bandgap ($\rm E_{g}$) energies. Interatomic distances,
binding and formation energies, and electronic bandgap energies are
given respectively in \AA, eV/atom, and eV.}
\vspace*{1cm}
\begin{center}
\begin{tabular}{ccccccccccc}
  \hline \hline
              &       $a$      & $d$(X-Y) & $d$(X-H) &  $d$(X-F)  &
 $d$(Y-H) & $d$(Y-F) & $\Delta z$ & $\rm E_B$ & $\rm E_{F}$  & $\rm E_{g}$ \\
 \hline
  $\alpha$-C$_{2}$         & 2.468 & 1.425 &      &      &      &    & 0    & -7.848 & &  0.0 \\
  C$_{2}$H$_{2}$  & 2.539 & 1.536 & 1.110 &      &      &   & 0.459 & -5.161 & -0.111& 3.47   \\
  C$_{2}$F$_{2}$  & 2.609 & 1.583 &       & 1.382 &      &  & 0.488 & -5.403 & -0.802 & 3.10  \\
 \hline
  $\alpha$-Si$_{2}$        & 3.897 & 2.250 &      &      &      &    & 0     & -3.894 &       & 0.0   \\
  $\beta$-Si$_{2}$         & 3.867 & 2.279 &      &      &      &    & 0.459 & -3.914 &       & 0.0   \\
  Si$_{2}$H$_{2}$ & 3.968 & 2.392 & 1.502 &      &      &    & 0.687 & -3.379 & -0.297 & 2.11  \\
  Si$_{2}$F$_{2}$ & 3.968 & 2.395 &       & 1.634 &      &   & \ 0.697 \ & \ -4.656 \ & \ -2.019 \ & 0.70  \\
 \hline
  $\alpha$-Ge$_{2}$        & 4.127 & 2.383 &      &      &      &    & 0     & -3.114 &       & 0.0  \\
  $\beta$-Ge$_{2}$         & 4.061 & 2.444 &      &      &      &    & 0.690 & -3.243 &       & 0.0  \\
  Ge$_{2}$H$_{2}$ & 4.091 & 2.473 & 1.563 &      &      &    & 0.730 & -2.882 & 0.107 & 0.95 \\
  Ge$_{2}$F$_{2}$ & 4.182 & 2.492 &       & 1.790 &      &   & 0.617 & -3.892 & -1.349 & 0.19  \\
 \hline
  $\alpha$-Sn$_{2}$        & 4.798 & 2.770 &      &      &      &    & 0     & -2.581 &       & 0.0 \\
  $\beta$-Sn$_{2}$         & 4.639 & 2.841 &      &      &      &    & 0.947 & -2.728 &       & 0.0 \\
  Sn$_{2}$H$_{2}$ & 4.719 & 2.846 & 1.738 &      &      &    & 0.824 & -2.517 & -0.030 & 0.45  \\
  Sn$_{2}$F$_{2}$ & 5.028 & 2.951 &       & 1.970   &      & & 0.531 & -3.625 & -1.581 & 0.0 \\
 \hline
  $\alpha$-SiC             & 3.100 & 1.790 &      &      &      &    & 0     & -5.905 & & 2.54   \\
  $\beta$-SiC              & 3.098 & 1.788 &      &      &      &    & 0.001 & -5.906 & & 2.54   \\
  SiCH$_{2}$      & 3.124 & 1.892 & 1.108 &       & 1.497 & & 0.573  & -4.366 & -0.288 & 4.04  \\
  SiCF$_{2}$ \      & \ 3.168 \ & \ 1.914 \ &       & 1.445 &      &1.609& 0.563& -5.096 & -1.463&  1.94  \\
 \hline \hline
\end{tabular}
\end{center}
\label{tab1}
\end{table}

\vfill
\eject
\pagebreak

\begin{figure}[ht]
\centering{
\includegraphics[width=100mm, angle=0.0]{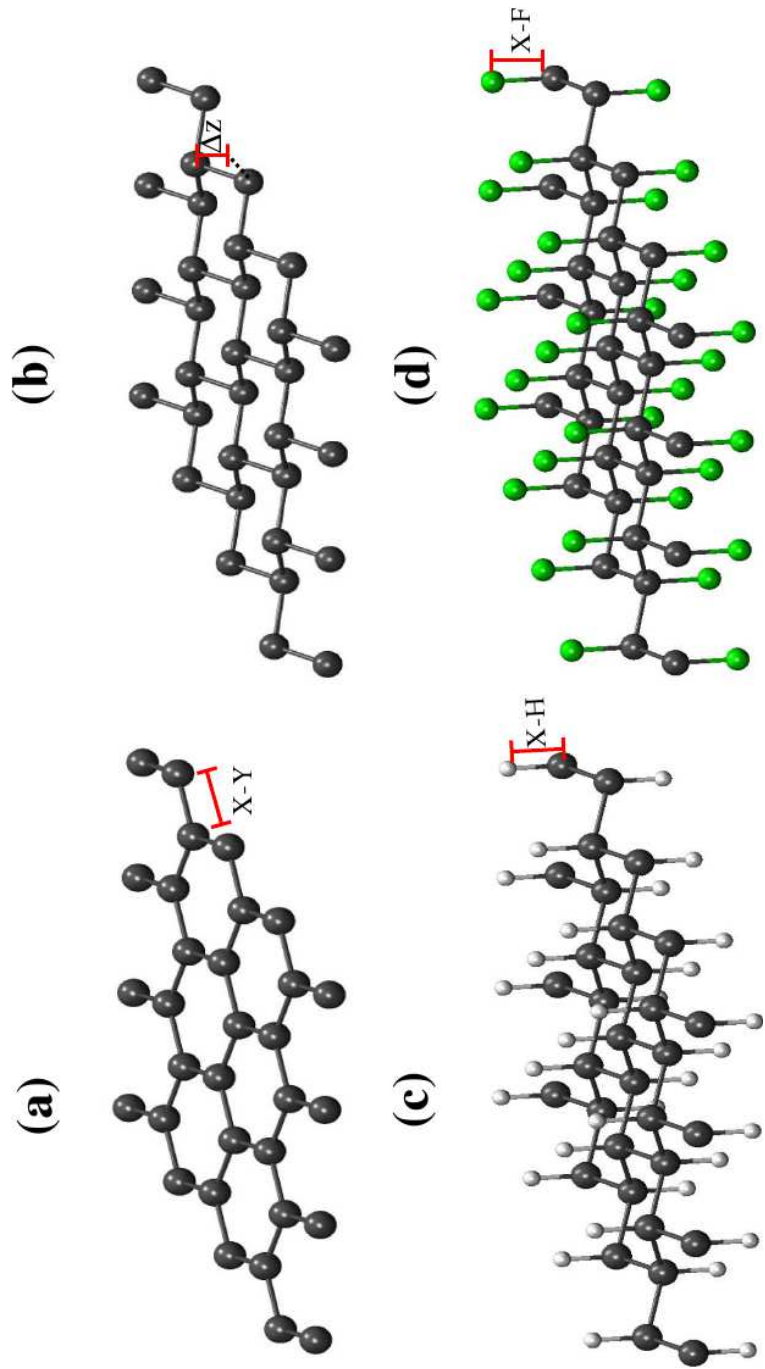}
\caption{(color online) Schematic representation of group-IV two dimensional materials:
(a) flat graphene-like ($\alpha$), (b) buckled graphene-like ($\beta$),
(c) hydrogenated graphane-like, and (d) fluorinated graphane-like
structures. The figure also indicate the interatomic distance labels, consistent
with the ones in \ref{tab1}. Black, grey and green spheres represent
group-IV, hydrogen and fluorine atoms, respectively.}
}
\label{fig1}
\end{figure}
\pagebreak

\begin{figure}[ht]
\centering{
\includegraphics[width=100mm, angle=0.0]{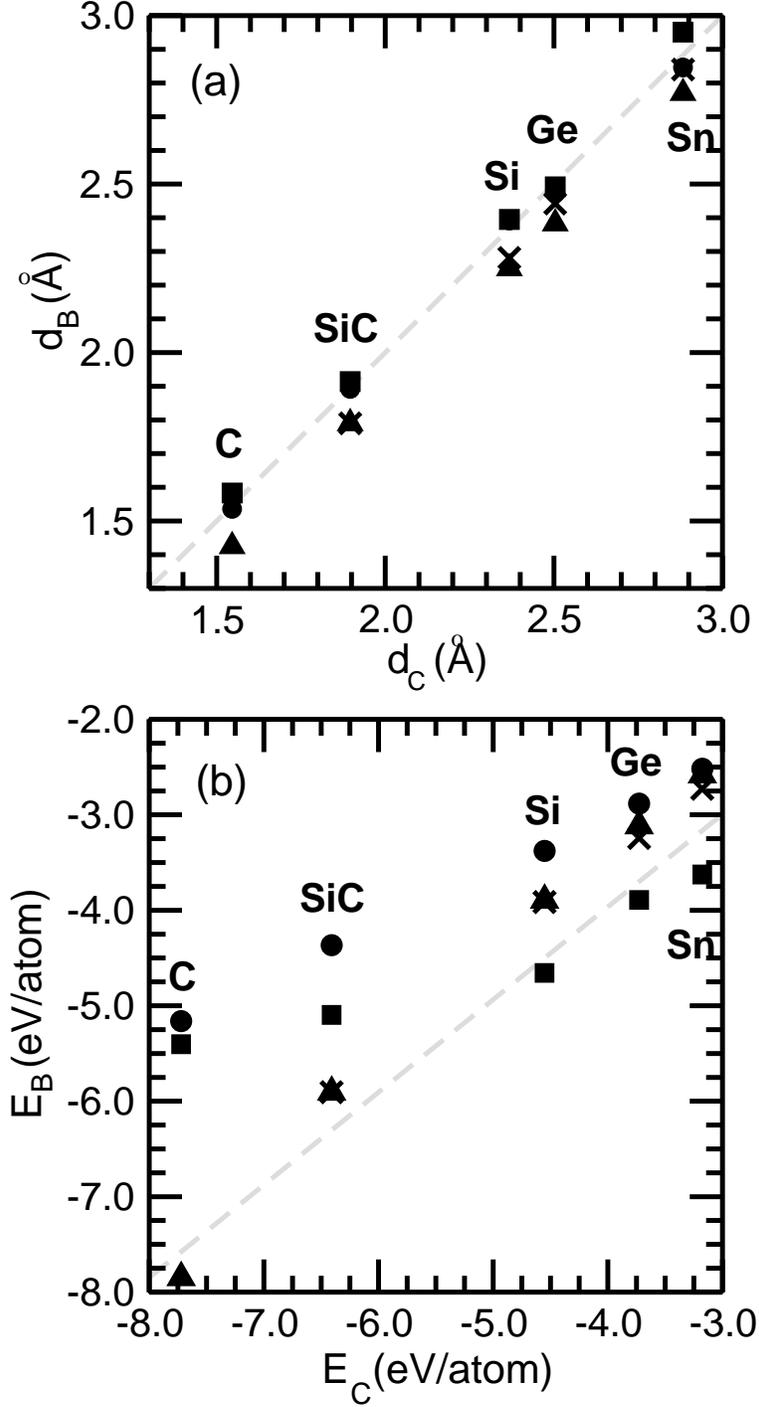}
\caption{Properties of group-IV (C, SiC, Si, Ge, and Sn)
graphene-like (in flat and buckled sheets) and graphane-like structures
(with full coverage of H or F atoms). The figure shows the (a) interatomic distances
($\rm d_B$) and  (b) and binding energies ($\rm E_B$) of those structures as
function of the respective distances ($\rm d_C$) and binding energies ($\rm E_C$)
in the diamond cubic (or zinc-blende) crystalline structures.
The $\blacktriangle$, $\times$, $\bullet$, and $\blacksquare$ symbols represent
respectively the flat graphene-like ($\alpha$), buckled graphene-like ($\beta$),
hydrogenated graphane-like, and fluorinated
graphane-like structures. The dashed lines are only a guide to the eye,
representing the properties $\rm d_B=d_C$ in (a) and  $\rm E_B=E_C$ in (b).}
}
\label{fig2}
\end{figure}
\pagebreak

\begin{figure}[ht]
\centering{
\includegraphics[width=140mm, angle=0.0]{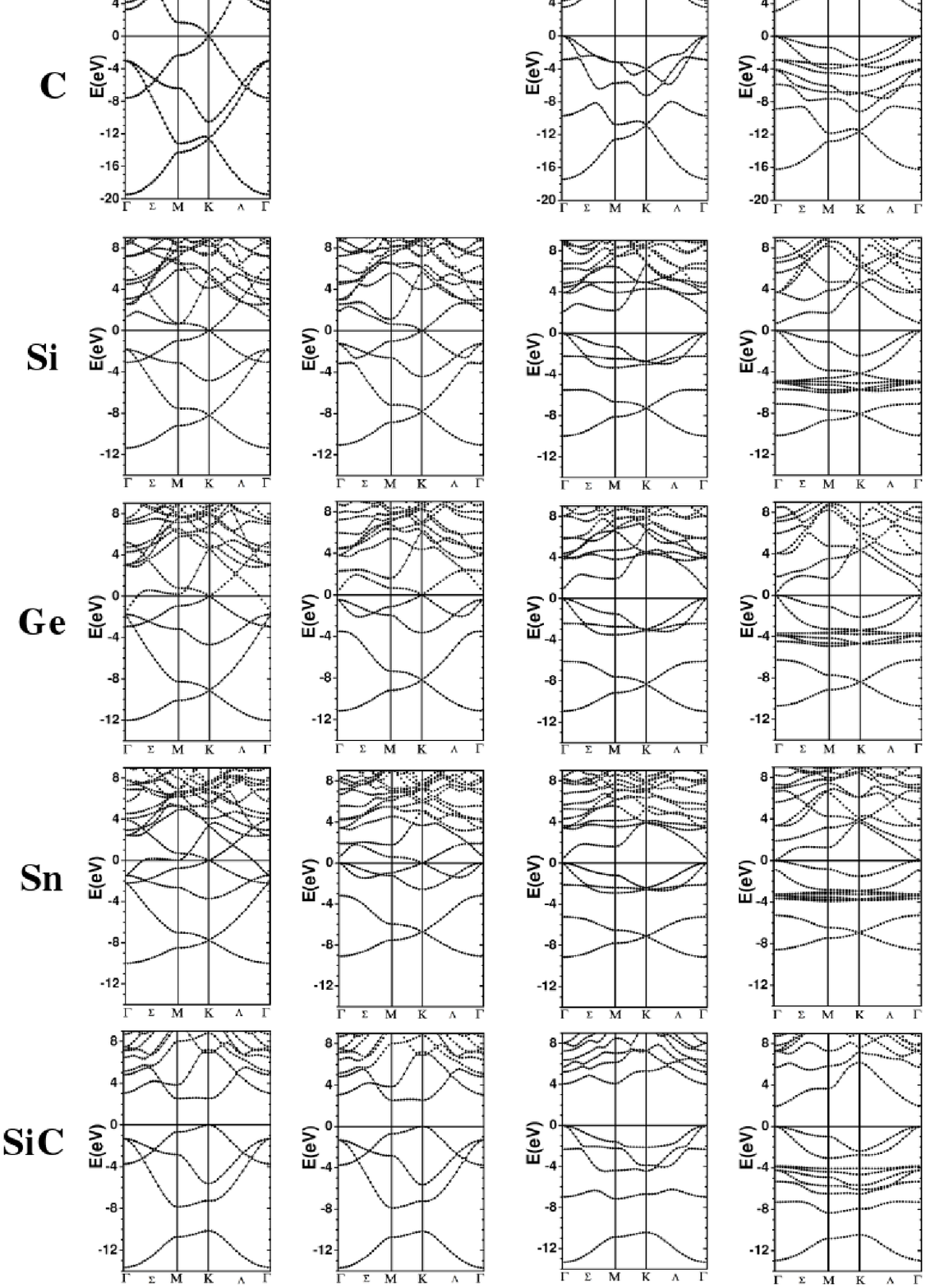}
%\vspace*{2cm}
\caption{Electronic band structure of group-IV in (a) flat graphene-like
($\alpha$), (b) buckled graphene-like ($\beta$), (c) hydrogenated graphane-like,
and (d) fluorinated graphane-like structures.}
}
\label{fig3}
\end{figure}
\pagebreak


\begin{thebibliography}{99}
\bibitem{solid} The properties of group-IV solids were computed considering
the diamond cubic (or zinc blende) stable structures within the same
approximations used in all the other calculations performed here.
\end{thebibliography}
\end{document}